\documentclass[12pt]{article}

\usepackage{amsmath,amssymb,cite}
\usepackage{cite}

\usepackage[unicode,linktocpage=true,plainpages=false,pdfpagelabels=false]{hyperref}

\newcommand{\pp}{\varphi}
\newcommand{\dd}{\partial}
\newcommand{\de}{\delta}
\newcommand{\m}{\mu}
\newcommand{\n}{\nu}
\newcommand{\ls}{\left(}
\newcommand{\rs}{\right)}
\newcommand{\al}{\alpha}
\newcommand{\be}{\beta}
\newcommand{\te}{\theta}
\newcommand{\ff}{\varphi}
\newcommand{\ta}{\tau}
\newcommand{\ga}{\gamma}
\newcommand{\str}[1]{\mathrel{\mathop{\longrightarrow}\limits_{#1}}}

\begin{document}

\title{Exact relation between canonical and metric energy-momentum tensors for higher derivative\\ tensor field theories}
\author{
R.~V.~Ilin\thanks{E-mail: st030779@student.spbu.ru},
S.~A.~Paston\thanks{E-mail: s.paston@spbu.ru}\\
{\it Saint Petersburg State University, Saint Petersburg, Russia}
}
\date{\vskip 15mm}
\maketitle

\begin{abstract}
We discuss the relation between canonical and metric energy-momentum tensors for field theories with actions that can depend on the higher derivatives of tensor fields in a flat spacetime. In order to obtain it we use a modification of the Noether's procedure for curved space-time.
For considered case the difference between these two tensors turns out to have more general form  than for theories with no more than first order derivatives. Despite this fact we prove that the difference between corresponding integrals of motion still has the form of integral over 2-dimensional surface that is infinitely remote in the spacelike directions.
\end{abstract}

\newpage

\section{Introduction}
Since the usage of symmetries which a field theory has is one the most useful tool in modern physics, the Noether's theorem \cite{noether} proven in the beginning of the XX century  plays a very significant role in it. This theorem states that each global continuous symmetry of the action corresponds to a conserving quantity. For example, spacetime translations correspond to so called canonical (it is also called "Noether") energy-momentum tensor (EMT). Its $00$-component which corresponds to the time translations is of particular importance as it defines energy density and hence has strong connection with energy conservation law. In the simplest case of scalar field theory with no more than 1st derivatives in Lagrangian it has the following form:
\begin{equation}
\mathcal{T}^0{}_0=\frac{\dd L}{\dd\dot{\phi}}\dot{\phi}-L,
\end{equation}
where $L$ is a Lagrangian density, $\phi$ is a tensor field and dot denotes time derivative.
Integral over $t=const$ surface of this component is equal to the Hamiltonian of theory, so Noether's theorem is closely connected with the Hamiltonian approach and the quantum theory. Other integrals that can be obtained from $0i$  components of canonical EMT are the densities of the system's total momentum.

Integrals of motion arising from Noether's theorem, e.g. energy or momentum,  are usually observable. In contrast, it is not always possible to interpret its densities (e.g. canonical EMT components) as observables. Probably the most infamous example of this issue is the inability to unambiguously define an energy density in the presence of gravity --- well-known problem of non-localizability of gravitational energy (see, for example \cite{faddeev-UFN1982}). \par
Another disadvantage of canonical EMT is that it is generally not symmetric, so there is no simple connection between EMT and angular momentum density tensor, see, for example, \cite{Weinberg}, \S7.4.

Another definition of EMT, arising from consideration of matter coupling with gravity in the framework of General Relativity (GR), turns out to be free from the problems mentioned above. This type of EMT is often called "metric" EMT (it is also called Hilbert or Belinfante EMT) because it can be obtained by varying the action w.r.t. spacetime metric $g_{\mu\nu}$:
\begin{equation}
T^{\mu\nu}=-\frac{2}{\sqrt{-g}}\frac{\delta S}{\delta g_{\mu\nu}}\Bigg|_{g_{\mu\nu}=\eta_{\mu\nu}},
\label{metric}
\end{equation}
where $S$ is a matter action with certain "minimal" coupling with gravity, $\eta_{\mu\nu}$ is a Minkowski metric. The metric EMT is automatically symmetric due to the symmetricity of metric, and its components are observable because they are equal to the r.h.s. of Einstein's equations and hence they define the curvature of a spacetime induced by matter.\par
For the first time the connection between canonical and metric EMT was noted by Belinfante \cite{Belifante,Belifante2} and Rosenfeld \cite{rosenfeld}. They showed that for field theories with first order derivative Lagrangians these two EMT definitions are equivalent in such a sense that they differ from each other by a divergent-free term:
\begin{equation}
\mathcal{T}^{\mu\nu}+\dd_{\alpha}B^{\alpha\mu\nu}=T^{\mu\nu},\qquad B^{\alpha\mu\nu}=-B^{\mu\alpha\nu}.
\label{eqviw}
\end{equation}
Such structure of the difference between metric and canonical EMT guarantees that both of them satisfy local conservation law:
\begin{equation}
\dd_\mu\mathcal{T}^{\mu\nu}=\dd_\mu T^{\mu\nu}=0.
\label{sp1}
\end{equation}
Moreover, the anti-symmetry of $B^{\alpha\mu\nu}$ leads to the fact that the integrals of motion  corresponding to $\mathcal{T}^{\mu\nu}$ and $T^{\mu\nu}$ are equal to each other up to  the integral
\begin{equation}\label{sp1.1}
\int\! d^3 x (T^{0\nu}-\mathcal{T}^{0\nu})=\int\! d^3x\, \dd_{\alpha}B^{\alpha0\nu}=\int\! d^3x\, \dd_{k}B^{k0\nu}
=\int\limits_W\! dS_k\, B^{k0\nu}
\end{equation}
(here $k=1,2,3$) over an infinitely remote $2$-dimensional surface $W$, which vanishes if the fields are decreasing fast enough.
It is worth noting that no comparison between canonical and metric EMT was made in the first Belinfante work \cite{Belifante}. Instead the main purpose of that work was to find such an addition to Noether's EMT that makes it symmetric and does not violate conservation law \eqref{sp1}. The further development of this approach is the method  \cite{GoMa} of construction of the EMT based on a covariant analogue of Noether's theorem.

To illustrate the relation \eqref{eqviw} one may consider the simple case of sourceless electrodynamics. For this theory canonical EMT takes the following form:
\begin{equation}
\mathcal{T}^{\mu\nu}=-F^{\mu\alpha}\dd^{\nu}A_{\alpha}+\frac{1}{4}\eta^{\mu\nu}F_{\alpha\beta}F^{\alpha\beta},
\label{sp2}
\end{equation}
where $A_{\alpha}$ is a vector potential and $F_{\alpha\beta}$ is an electromagnetic field tensor. The EMT \eqref{sp2} is not symmetric; however, the addition of the expression
$ F^{\mu\alpha}\dd_{\alpha}A^{\nu},$
which is on-shell equivalent to the addition of $\dd_{\alpha}B^{\alpha\mu\nu}$, makes \eqref{sp2} symmetric and equal to the metric EMT $T^{\mu\nu}$.

One may ask whether the relation between canonical and metric EMT \eqref{eqviw} or its generalization holds true for theories more general than those with first order derivative Lagrangians. The case of the second order derivative Lagrangian was considered, for example, in \cite{Ohanian} (see also \cite{Mickevich1969}).
The present paper is devoted to investigation of the most general case of Poincar\`e invariant theory of tensor fields with Lagrangian density which may depend on field derivative of arbitrary order. We show that for such theory the relation between canonical and metric EMT still takes the form of \eqref{eqviw} but $B^{\alpha\mu\nu}$ may not be anti-symmetric. However, the difference \eqref{sp1.1} between the corresponding integrals of motion is a surface term as before but it has more complex structure than in \eqref{sp1.1}.

\section{Search for the relation between canonical and metric EMT}
Consider Poincar\`e invariant theory of tensor field $\pp_A$ (we use multi-index $A\equiv\mu_1\mu_2...\mu_m$ for brevity, in this terms $\pp_A$ is tensor of rank $m$) in the flat spacetime with Lagrangian density depending on the fields and their derivatives up to $n$th order:
\begin{equation}
L(\pp_A, \dd_{\tau_1}\pp_A,\ldots,\dd_{\tau_1}...\dd_{\tau_n}\pp_A).
\label{lagr}
\end{equation}
It is easy to show that equations of motion for this theory take the form:
\begin{equation}\label{sp3}
\frac{\partial{L}}{\partial\varphi_A}+
\sum_{j=1}^{n}(-1)^j\partial_{\tau_1}...\partial_{\tau_j}
\frac{\partial{L}}{\partial(\partial_{\tau_1}...\partial_{\tau_j}\varphi_A)}=0.
\end{equation}
According to Noether's theorem one can find the EMT expression for this theory. Probably the simplest way to do it is to consider a derivative of the Lagrangian density $\dd_\theta L$ and note that it can be rewritten using the chain rule:
\begin{equation}\label{sp4}
\partial_{\theta}{L}=\frac{\partial{L}}{\partial\varphi_A}\partial_{\theta}\varphi_A+
\sum_{j=1}^{n}
\frac{\partial{L}}{\partial(\partial_{\tau_1}...\partial_{\tau_j}\varphi_A)}\partial_{\tau_1}...\partial_{\tau_j}\partial_{\theta}\varphi_A.
\end{equation}
Then one need to apply the product rule in \eqref{sp4} to make use of equations of motion \eqref{sp3}. As a result one can obtain the following local conservation law:
\begin{equation}\label{sp4.1}
\dd_\tau\mathcal{T}^\tau{}_\te=0,
\end{equation}
where the expression
\begin{equation}\label{sp5}
\mathcal{T}^\tau{}_\te=\sum_{j=1}^{n}\sum_{i=1}^{j}(-1)^{i+j}
\left(\partial_{\tau_{i+1}}...\partial_{\tau_{j}}\frac{\partial{L}}{\partial(\partial_{\tau}\partial_{\tau_2}...\partial_{\tau_j}\varphi_A)}\right)
\partial_{\tau_2}...\partial_{\tau_i}\partial_{\theta}\varphi_A-{L}\delta^{\tau}_{\theta}.
\end{equation}
defines the canonical EMT for theory \eqref{lagr}.
Hereinafter we assume for brevity that $\partial_{\tau_i}...\partial_{\tau_j}$ means $\partial_{\tau_i}$ for $j=i$ and factor $1$ for $j=i-1$.

For further consideration it is convenient to use \emph{the variation of the form of a function} (VFF) with respect to the infinitesimal coordinate transformation:
\begin{equation}\label{sp6}
x^{\mu}\rightarrow x'^{\mu}= x^{\mu}+\varepsilon^{\mu}(x).
\end{equation}
For a field $\psi(x)$ we define such variation by the following formula:
\begin{equation}\label{sp6.1}
\bar{\delta}\psi=\psi'(x)-\psi(x).
\end{equation}
It should be noted that VFF for tensors are generally covariant (i.e. tensors) since  differences between tensors at the same point are considered here. If the transformation law for field $\psi(x)$ with respect to \eqref{sp6} is known then VFF for it is uniquely defined. For the covariant tensor field $\pp_A(x)$ (we recall that $A\equiv\mu_1\mu_2...\mu_m$) form variation has the following form:
\begin{equation}
\bar{\delta}\varphi_A=-\xi^{\alpha}\partial_{\alpha}\varphi_A-(\partial_{\mu_1}\xi^{\alpha})\varphi_{\alpha\mu_2...\mu_m}-
(\partial_{\mu_2}\xi^{\alpha})\varphi_{\mu_1\alpha\mu_3...\mu_m}-\ldots .
\label{var}
\end{equation}
It's easy to see that VFF defined by \eqref{sp6.1} gives the same (up to a sign) result for tensors as the definition of Lie derivative (see, for example \cite{Trautman2008}). \par
The VFF for a spacetime metric $g_{\mu\nu}$ and its determinant $g$ can be easily calculated:
\begin{align}
\bar{\delta}g_{\m\n}&=-D_{\m}\xi_{\n}-D_{\n}\xi_{\m},\label{Variation gmu}\\
&\bar{\delta}g=-2\left(D_{\alpha}\xi^{\alpha}\right)g.
\label{Variation g}
\end{align}
One also can obtain the expression for VFF of any scalar density, i.e. the quantity with the same transformation law as for $\sqrt{-g}$. For example, for the arbitrary Lagrangian density $\hat L$ we have:
\begin{equation}
\bar{\delta}\hat L=-\partial_{\alpha}(\xi^{\alpha}\hat L).
\label{ScalarDensity}
\end{equation}

Now we move directly to establishing the relation between canonical \eqref{sp5} and metric \eqref{metric} EMT. In order to give the concrete meaning to the latter we should include a "minimal coupling"{} with gravity to the theory \eqref{lagr} (for higher derivative theories this coupling it not always unique, see discussion in Sec. 3). To include minimal coupling one needs to replace all partial derivatives by covariant ones, flat metric $\eta_{\mu\nu}$ by $g_{\mu\nu}$ in all contractions and finally one adds a factor $\sqrt{-g}$ in order to make integration measure invariant under diffeomorphisms. New Lagrangian density which has been obtained through this procedure depends not only on $\ff_A$ and its derivatives  up to $n$th order but also on spacetime metric and its derivatives:
\begin{equation}\label{sp7}
\hat L(\pp_A, \dd_{\tau_1}\pp_A,\ldots,\dd_{\tau_1}...\dd_{\tau_n}\pp_A,g_{\mu\nu}, \dd_{\tau_1}g_{\mu\nu},\ldots,\dd_{\tau_1}...\dd_{\tau_n}g_{\mu\nu}).
\end{equation}\par

On the one hand, VFF for $\hat L$ is given by \eqref{ScalarDensity}. On the other hand,  since VFF has the properties of differential operator, one can express it in terms of VFF for arguments of \eqref{sp7} using the chain rule:
\begin{equation}\label{sp8}
\bar{\delta} \hat L=\sum_{j=0}^n \frac{\partial \hat L}{\partial(\partial_{\tau_1}...\partial_{\tau_j}\varphi_A)}\partial_{\tau_1}...\partial_{\tau_j}\bar{\delta}\varphi_A+
\sum_{j=0}^n \frac{\partial \hat L}{\partial(\partial_{\tau_1}...\partial_{\tau_j}g_{\mu\nu})}\partial_{\tau_1}...\partial_{\tau_j}\bar{\delta}g_{\mu\nu}.
\end{equation}
Using the equations of motion \eqref{sp7} (which are \eqref{sp3} where $L$ is replaced by $\hat L$) and the relation
\begin{equation}
\frac{\partial \hat L}{\partial g_{\mu\nu}}+
\sum_{j=1}^n (-1)^j\partial_{\tau_1}...\partial_{\tau_j}
\frac{\partial \hat L}{\partial(\partial_{\tau_1}...\partial_{\tau_j}g_{\mu\nu})}=-\frac{1}{2}\sqrt{-g}\hat T^{\mu\nu},
\label{Metric}
\end{equation}
that follows from the definition of metric EMT $\hat T^{\mu\nu}$ for theory \eqref{sp7},  one can replace the terms with $j=0$ in the first and second sum in \eqref{sp8}.
As a result equation \eqref{sp8} takes the form:
\begin{equation}\label{sp9}
\begin{split}
\bar{\delta}\hat L=&\sum_{j=1}^n\left[(-1)^{j+1}\left(\partial_{\tau_1}...\partial_{\tau_j}
\frac{\partial\hat L}{\partial(\partial_{\tau_1}...\partial_{\tau_j}\varphi_A)}\right)\bar{\delta}\varphi_A+
\frac{\partial \hat L}{\partial(\partial_{\tau_1}...\partial_{\tau_j}\varphi_A)}\partial_{\tau_1}...\partial_{\tau_j}\bar{\delta}\varphi_A\right]+\\
+&\sum_{j=1}^n \left[(-1)^{j+1}\left(\partial_{\tau_1}...\partial_{\tau_j}
\frac{\partial \hat L}{\partial(\partial_{\tau_1}...\partial_{\tau_j}g_{\mu\nu})}\right)\bar{\delta}g_{\mu\nu}+
\frac{\partial \hat L}{\partial(\partial_{\tau_1}...\partial_{\tau_j}g_{\mu\nu})}\partial_{\tau_1}...\partial_{\tau_j}\bar{\delta}g_{\mu\nu}\right]-\\-
&\frac{1}{2}\sqrt{-g}\hat T^{\mu\nu}\bar{\delta}g_{\mu\nu}.
\end{split}
\end{equation}
Taking the left-hand side of this equation in the form \eqref{ScalarDensity} and using product rule (analogously to the procedure described above \eqref{sp5}) one can obtain the following on-shell relation
\begin{equation}\label{sp10}
\begin{split}
\partial_{\tau}\Biggl[\sum_{j=1}^{n}\sum_{i=1}^{j}(-1)^{i+j}
\left(\partial_{\tau_{i+1}}...\partial_{\tau_{j}}\frac{\partial \hat L}{\partial(\partial_{\tau}\partial_{\tau_2}...\partial_{\tau_j}\varphi_{A})}\right)
\partial_{\tau_2}...\partial_{\tau_i}\bar{\delta}\varphi_A+&\\
+\sum_{j=1}^{n}\sum_{i=1}^{j}(-1)^{i+j}
\left(\partial_{\tau_{i+1}}...\partial_{\tau_{j}}\frac{\partial \hat L}{\partial(\partial_{\tau}\partial_{\tau_2}...\partial_{\tau_j}g_{\mu\nu})}\right)&
\partial_{\tau_2}...\partial_{\tau_i}\bar{\delta}g_{\mu\nu}\Biggr]+\\+
\partial_{\alpha}(\xi^{\alpha} \hat L)&-\frac{1}{2}\sqrt{-g}\hat T^{\mu\nu}\bar{\delta}g_{\mu\nu}=0.
\end{split}
\end{equation}
Now let us substitute \eqref{var} and \eqref{Variation gmu} for $\bar{\delta}\varphi_A$ and $\bar{\delta}g_{\mu\nu}$ into it and then set the spacetime metric equal to flat one $g_{\m\n}\to\eta_{\m\n}$. Note that in this limit $\hat L$ becomes equal to $L$ and $\hat T^{\mu\nu}$ becomes equal to $T^{\mu\nu}$ (see \eqref{metric}) and hence for any $\xi^\m(x)$ one has the following equation:
\begin{equation}
\begin{split}
\partial_{\tau}\Biggl\{\sum_{j=1}^{n}\sum_{i=1}^{j}(-1)^{i+j}\Biggl[
\left(\partial_{\tau_{i+1}}...\partial_{\tau_{j}}\frac{\partial L}{\partial(\partial_{\tau}\partial_{\tau_2}...\partial_{\tau_j}\varphi_{A})}\right)
\partial_{\tau_2}...\partial_{\tau_i}\Bigl(\xi^{\alpha}\partial_{\alpha}\varphi_A+(\partial_{\mu_1}\xi^{\alpha})\varphi_{\alpha\mu_2...\mu_m}+\ldots\Bigr)+\\
+2\left(\partial_{\tau_{i+1}}...\partial_{\tau_{j}}\frac{\partial \hat L}{\partial(\partial_{\tau}\partial_{\tau_2}...\partial_{\tau_j}g_{\mu\nu})}\right)
\Bigg|_{g_{\ga\de}=\eta_{\ga\de}}
\partial_{\tau_2}...\partial_{\tau_i}\dd_{\mu}\xi_{\nu}\Biggr]\Biggr\}-\partial_{\alpha}(\xi^{\alpha} L)-T^{\mu\nu}\dd_{\mu}\xi_{\nu}=0
\label{Riemanneq}
\end{split}
\end{equation}
(again, we recall that $A\equiv\mu_1\mu_2...\mu_m$).
As the vector $\xi^{\al}$ and its derivatives at one point can be chosen independently from each other the equation \eqref{Riemanneq} is equivalent to a system of $n+2$ equations which can be derived from setting to zero the coefficients of $\xi^{\al}$ and its derivatives (after a symmetrization by the indices that are contracted with derivatives indices).\par

Now we analyze the system of equations obtained. It's easy to show that by setting to zero the coefficient of $\xi^{\al}$ without derivatives one can obtain the local conservation law \eqref{sp4.1} for canonical EMT \eqref{sp5}. This is not surprising because the reasoning made when deriving \eqref{Riemanneq} in particular case of $\xi^{\al}=const$ is the same as it was made for deriving canonical EMT \eqref{sp5}. As a further step we consider the coefficient of $\dd_{\be}\xi_{\al}$. It is not difficult to prove that setting this coefficient to zero leads to the following equation:
\begin{equation}\label{sp11}
T^{\be\alpha}-\mathcal{T}^{\be\alpha}=\dd_{\ta}B^{\ta\be\al},
\end{equation}
where
\begin{equation}
\begin{split}
B^{\ta\be\al}&=\sum_{j=1}^{n}\Biggl[\sum_{i=1}^{j}(-1)^{i+j}
\left(\partial_{\tau_{i+1}}...\partial_{\tau_{j}}\frac{\partial L}{\partial(\partial_{\tau}\partial_{\tau_2}...\partial_{\tau_j}\varphi_A)}\right)
\Bigl((i-1)\delta^{\be}_{\tau_2}\partial_{\tau_3}...\partial_{\tau_i}\partial^{\alpha}\varphi_A+\\
&+\delta_{\mu_1}^{\be}\partial_{\tau_2}...\partial_{\tau_i}\varphi_{\alpha\mu_2...\mu_m}+\ldots\Bigr)+
2(-1)^{j+1}\left(\partial_{\tau_{2}}...\partial_{\tau_{j}}
\frac{\partial \hat L}{\partial(\partial_{\tau}\partial_{\tau_2}...\partial_{\tau_j}g_{\be\al})}\right)\Bigg|_{g_{\ga\de}=\eta_{\ga\de}}
\Biggr]=0.
\label{2nd}
\end{split}
\end{equation}
The equation \eqref{sp11} reproduces the formula \eqref{eqviw} and give the search relation between canonical and metric EMT. At this point, \eqref{2nd} gives the exact expression for the difference between these two tensors in terms of derivatives of Lagrangian density $L$. However, there is no evidence of this difference $B^{\ta\be\al}$ being anti-symmetric as it was for the first order derivative Lagrangian. Nevertheless, this term is divergence-free because of the fact that the divergence of $\mathcal{T}^{\be\al}$ and $T^{\be\al}$ is equal to zero which follows from Noether's theorem and well-known local covariant conservation law in the flat limit $g_{\m\n}\to\eta_{\m\n}$ which is satisfied on-shell respectively.
To the contrary, the difference between corresponding to $\mathcal{T}^{\be\al}$ and $T^{\be\al}$ integrals of motion is no more automatically equal to surface term (as it was for theory with first order Lagrangian, see \eqref{sp1.1}). To ensure that it is equal to surface term one have to analyze the rest of the equations from system \eqref{Riemanneq} arising from setting to zero coefficients of second and higher order derivatives of $\xi^{\al}$.\par
To show it, let us write \eqref{Riemanneq} in the form
\begin{equation}\label{sp12}
\dd_\ta\left[\sum_{l=0}^{n}R_{(l)}^{\ta\al_1...\al_l\al}\dd_{\al_1}...\dd_{\al_l}\xi_\al\right]=T^{\mu\nu}\dd_{\mu}\xi_{\nu},
\end{equation}
where $R_{(l)}^{\ta\al_1...\al_l\al}$ (we assume that they are symmetric over indices $\al_1...\al_l$) can be calculated by comparing \eqref{sp12} and \eqref{Riemanneq}. As the coefficients of $\xi^{\al}$ and $\dd_{\be}\xi_{\al}$ were analyzed before, it is obvious that
\begin{equation}\label{sp13}
R_{(0)}^{\ta\al}=\mathcal{T}^{\ta\al},\qquad
R_{(1)}^{\ta\al_1\al}=B^{\ta\al_1\al},
\end{equation}
where $B^{\ta\al_1\al}$ is from \eqref{2nd}. It appears that exact expressions for $R_{(l)}^{\ta\al_1...\al_l\al}$ with $l\ge2$ are unnecessary for further considerations, so they are given in Appendix.\par
By taking derivative of the equation \eqref{sp12}, it can be written in the following form:
\begin{equation}\label{sp14}
\sum_{l=0}^{n}\ls
(\dd_\ta R_{(l)}^{\ta\al_1...\al_l\al})\dd_{\al_1}...\dd_{\al_l}\xi_\al+
R_{(l)}^{\ta\al_1...\al_l\al}\dd_\ta\dd_{\al_1}...\dd_{\al_l}\xi_\al\rs
=T^{\mu\nu}\dd_{\mu}\xi_{\nu}.
\end{equation}
By setting to zero coefficient of maximal order of  $(n+1)$th derivative of $\xi_{\al}$ one can obtain the following equation:
\begin{equation}\label{sp15}
R_{(n)}^{(\ta\al_1...\al_n)\al}=0.
\end{equation}
This procedure can be applied to other coefficients, in particular to $l$-th order derivative of $\xi_{\al}$ with $2\le l\le n$ which allow us to write recursive formula:
\begin{equation}\label{sp16}
\dd_\ta R_{(l)}^{\ta\al_1...\al_l\al}+
R_{(l-1)}^{(\al_1...\al_l)\al}=0.
\end{equation}
It's worth noting that two equations for $l=1$ and $l=0$ are the abovementioned equations \eqref{sp11} and \eqref{sp4.1} (with taking into account \eqref{sp13}) respectively. In formulas \eqref{sp15} and \eqref{sp16} indices in parenthesis mean fully symmetric part of the tensor over them.\par
It follows from \eqref{sp16} that vector $\dd_\ta\dd_{\al_1}...\dd_{\al_l}R_{(l)}^{\ta\al_1...\al_l\al}$ does not (up to a sign) depend on $l$ at $1\le l\le n$ and taking \eqref{sp15} into account one can see that it is equal to zero. Then one may consider this vector for $l=1$ and by taking into account \eqref{sp13} obtain that
\begin{equation}\label{sp17}
\dd_\ta\dd_{\al_1}B^{\ta\al_1\al}=0.
\end{equation}
This relation verifies our previous discussion (see notes after \eqref{2nd}) that the difference \eqref{sp11} between canonical and metric EMT is divergence-free.\par
Now we can write down the equation \eqref{sp16} with $\al_1=...=\al_l=0$ (symmetrization can be omitted here):
\begin{equation}\label{sp18}
\dd_k R_{(l)}^{k0...0\al}+\dd_0 R_{(l)}^{00...0\al}+R_{(l-1)}^{0...0\al}=0.
\end{equation}
Here we separated sum over repeated index $\ta$ into sum over spatial indices $k$ and term with $\ta=0$. Note that \eqref{sp15} with $\tau=\al_1=...=\al_l=0$ leads to the relation $R_{(n)}^{00...0\al}=0$. Consistently applying $(-\dd_0)^{l-2}$ to \eqref{sp18}, summing the result over $2\le l \le n$ and taking into account that $B^{\ta\al_1\al}=R_{(1)}^{\ta\al_1\al}$ and $R_{(n)}^{00...0\al}=0$ one can show that
\begin{equation}\label{sp19}
B^{00\al}=-\dd_k \sum_{l=2}^n (-\dd_0)^{l-2}R_{(l)}^{k0...0\al}.
\end{equation}
This result allows us to generalize the relation \eqref{sp1.1} for difference between canonical and metric EMT in the following form:
\begin{equation}\label{sp20}
\begin{split}
\int\! d^3 x (T^{0\nu}-\mathcal{T}^{0\nu})=\int\! d^3x\, \dd_{\alpha}B^{\alpha0\nu}=
\int\! d^3x\,& \dd_{k}\ls B^{k0\nu}+\sum_{l=2}^n (-\dd_0)^{l-1}R_{(l)}^{k0...0\nu}\rs=\\&
=\int\limits_W\! dS_k\, \ls B^{k0\nu}+\sum_{l=2}^n (-\dd_0)^{l-1}R_{(l)}^{k0...0\nu} \rs.
\end{split}
\end{equation}
To sum up, despite $B^{\ta\be\al}$ in general case is not anti-symmetric over first 2 indices, the difference between corresponding to canonical and metric EMT integrals of motion can be presented as the surface integral over surface $W$ infinitely remote in spacelike directions and vanishes for matter fields with proper decreasing asymptotics.\par

\section{Conclusion}
For arbitrary Poincar\`e invariant theory of tensor fields with high order derivative Lagrangian there is the relation \eqref{sp11} between canonical and metric EMT:
\begin{equation}\label{spz1}
T^{\be\alpha}-\mathcal{T}^{\be\alpha}=\dd_{\ta}B^{\ta\be\al},
\end{equation}
where $B^{\ta\be\al}$ is from \eqref{2nd}. In comparison to theories with first order derivative Lagrangians $B^{\ta\be\al}$ is no longer anti-symmetric over first two indices in case of higher derivatives. Instead, a weaker relation is satisfied:
\begin{equation}\label{spz2}
\dd_\ta\dd_{\be}B^{\ta\be\al}=0,\qquad
B^{00\al}=\dd_k \xi^k.
\end{equation}
Still, it appears that these equations are enough
in order to the divergence of metric and canonical EMT be equal to zero as well as the corresponding integrals of motion be equal to each other up to the surface terms.

We remind the reader that integrals of motion  corresponding to EMT  are the components of energy-momentum vector. As there are two EMT under considerations, one can introduce either "metric" or "canonical" vector:
\begin{equation}\label{spz3}
P^\al=\int\! d^3 x\, T^{0\al},\qquad
\mathcal{P}^\al=\int\! d^3 x\, \mathcal{T}^{0\al}.
\end{equation}
The conditions for the conserving of these quantities are the following expressions for field asymptotics:
\begin{equation}\label{spz4}
r^2 T^{k\al},r^2 \mathcal{T}^{k\al}\str{r\to\infty}0,
\end{equation}
where $r$ is distance in spatial direction.
According to \eqref{sp20} the difference between $P^\al$ and $\mathcal{P}^\al$ can be written in the form of surface integral
\begin{equation}\label{spz5}
P^\al-\mathcal{P}^\al=
\int\! d^3 x (T^{0\al}-\mathcal{T}^{0\al})=
\int\limits_W\! dS_k\, \sum_{l=1}^n (-\dd_0)^{l-1}R_{(l)}^{k0...0\al}
\end{equation}
(here we used \eqref{sp13}).
Whether its right hand side equal to zero or not depends on the asymptotic behavior of $\dd_0^{l-1}R_{(l)}^{k0...0\al}$ in the spatial directions. One can find the exact expressions for these quantities in terms of fields and their derivatives in the Appendix (formula \eqref{pril1}).

Note that the quantity $B^{\ta\be\al}$ \eqref{2nd} which defines the difference \eqref{spz1} contains the derivatives of Lagrangian density $\hat L$ with respect to spacetime derivatives of metric and hence it depends on the method of coupling of matter with gravity. For theories with no more than first order derivatives in Lagrangian this procedure (it was described before \eqref{sp7}) is well-defined and the expression for $B^{\ta\be\al}$ can always be written in terms of only derivatives of the Lagrangian density $L$ in the flat spacetime. For example, it's easy to show that for free vector field $\pp_\m$ in this case
\begin{equation}\label{spz6}
\frac{\dd\hat L}{\dd(\dd_{\ta}g_{\be\al})}=\frac{q^{\be\al\ta}+q^{\al\be\ta}-q^{\ta\be\al}-q^{\ta\al\be}-q^{\be\ta\al}-q^{\al\ta\be}}{4},
\qquad
q^{\ta\be\al}=\frac{\dd L}{\dd(\dd_\ta\pp_\be)}\pp^\al.
\end{equation}
By substituting this formula into \eqref{2nd}, one can obtain the well-known formula (see \cite{Belifante}):
\begin{equation}\label{spz7}
B^{\ta\be\al}=\frac{1}{2}\ls S^{\ta\be\al}+S^{\be\al\ta}+S^{\al\be\ta}\rs,
\end{equation}
where
\begin{equation}\label{spz8}
S^{\ta\be\al}=\frac{\dd L}{\dd\dd_{\ta}\pp_\be}\pp^\al-\frac{\dd L}{\dd\dd_{\ta}\pp_\al}\pp^\be
\end{equation}
is usually called spin tensor.

However, if one tries to minimally couple matter with gravity in theory with higher order derivative Lagrangian in the way it was described before \eqref{sp7}, one will find that it no longer  gives an unambiguous result for metric EMT \eqref{metric} and it will depend on the coupling scheme. The reason for this ambiguity lies in the different possible variants of ordering the covariant derivatives which arise in the Lagrangian density. This fact can be illustrated by an example of contribution for vector field Lagrangian density of the following form:
\begin{equation}
\Delta L=\pp^\mu(\dd^\nu \pp^\al)\dd_\mu\dd_\nu \pp_\al.
\end{equation}
By using the minimal coupling recipe from the discussion before \eqref{sp7} one can obtain two different expressions for contributions in $\Delta \hat{L}$ and there are no criteria to point out which one of them is more "minimal" than the other one:
\begin{equation}
\begin{split}
&\Delta \hat L_1=\sqrt{-g}g^{\ga\m}g^{\de\n}g^{\be\al}\pp_\ga(D_\de \pp_\be)D_\mu D_\nu \pp_\al
,\\
&\Delta \hat L_2=\sqrt{-g}g^{\ga\m}g^{\de\n}g^{\be\al}\pp_\ga(D_\de \pp_\be)D_\n D_\m \pp_\al.
\end{split}
\end{equation}
However, the difference between these two expressions which can be performed as
\begin{equation}
\Delta \hat L_1-\Delta \hat L_2=
\sqrt{-g}\pp_\ga(D_\de \pp_\be) R^{\be\al\ga\de}\pp_\al,
\end{equation}
already does not look "minimal" since it contains the Riemann curvature tensor $R^{\be\al\ga\de}$ and it's not difficult to prove that its contribution to the metric EMT is not equal to zero. Hence this "minimal" coupling procedure can be ambiguous.\par
The question about the exact expression for the difference between metric and canonical EMT can arise for example in the investigations of the gravitational energy definition problem in the framework of modified gravity.
For formulation of gravity in approach of embedding theory \cite{regge,deser,statja18,statja24} this problem was studied in \cite{statja46}. In this work it was shown that for formulation of gravity in the framework of splitting theory (version of embedding theory) that is theory with higher order derivative Lagrangian in the flat ambient spacetime \cite{statja25} the difference between metric and canonical EMT are not trivial. Furthermore it appears that metric EMT is non-trivial for Einstein solutions (which correspond to the GR solutions) while canonical EMT definition is zero for them.
A move from the canonical energy density definition, which is closely connected with Hamiltonian approach and hence quantum theory to observable metric energy density definition can be also useful in studies that highly involves situations where the energy density can play important role due to the quantum effects, for example, in Casimir effect investigations.

{\bf Acknowledgements.}
The work of one of the authors (R.~V.~Ilin) was supported by RFBR grant N~18-31-00169.

\section*{Appendix}
Here we give an exact expression for abovementioned (see \eqref{sp12}) quantities $R_{(l)}^{\ta\al_1...\al_l\al}$ with $l\ge 1$. To derive them one should take the coefficient of $\dd_{\al_1}...\dd_{\al_l}\xi_\al$ in the expression in the curly brackets in formula \eqref{Riemanneq}. The direct calculation shows that it can be written in the following form:
\begin{align}
\begin{split}
\tilde R_{(l)}^{\ta\al_1...\al_l\al}&=\sum_{j=1}^{n}\sum_{i=1}^{j}(-1)^{i+j}\left(\partial_{\tau_{i+1}}...\partial_{\tau_{j}}\frac{\partial L}{\partial(\partial_{\ta}\dd_{\al_2}...\dd_{\al_l}\dd_{\ta_{l+1}}...\partial_{\ta_j}\varphi_A)}\right)
\times\\ \times&
\left(C^l_{i-1}\de^{\al_1}_{\ta_{l+1}}\dd_{\ta_{l+2}}...\dd_{\ta_i}\dd^{\al}\pp_A+
C^{l-1}_{i-1}\dd_{\ta_{l+1}}...\dd_{\ta_i}\left(
\de^{\al_1}_{\mu_1}\pp^{\al}{}_{\mu_2\ldots\mu_m}+\de^{\al_1}_{\mu_2}\pp_{\mu_1}{}^{\al}{}_{\mu_3\ldots\mu_m}+
\ldots\right)\right)+
\\ +
2&\sum_{j=1}^n(-1)^{j+l}
\left(\partial_{\ta_{l+1}}...\partial_{\ta_j}\frac{\partial \hat L}{\partial(\partial_{\ta}\dd_{\al_2}...\dd_{\al_l}\dd_{\ta_{l+1}}...\partial_{\ta_j}g_{\al_1\al})}\right)
\Bigg|_{g_{\ga\de}=\eta_{\ga\de}},
\label{pril1}
\end{split}
\end{align}
where $C^{k}_{i}$ are binomial coefficients vanishing for $k>i$. Since $R_{(l)}^{\ta\al_1...\al_l\al}$ is assumed to be symmetric over $\al_1...\al_l$ by definition (see after \eqref{sp12}), they can be easily calculated by taking fully symmetric part of \eqref{pril1} over these indices. For brevity in \eqref{pril1} we assume that $A\equiv\mu_1\mu_2...\mu_m$. We also assume that $\partial_{\tau_i}...\partial_{\tau_j}$ denotes factor $1$ for $j=i-1$, factor $0$ for $j<i-1$ and $\partial_{\tau_i}$ for $j=i$.
When deriving \eqref{pril1} we have used simple formula:
\begin{equation}
\begin{split}
G^{\tau_1...\tau_i}\partial_{\tau_1}...\partial_{\tau_i}(fg)=
G^{\tau_1...\tau_i}\sum_{l=0}^{i}C^l_i(\partial_{\tau_1}...\partial_{\tau_l}f)(\partial_{\tau_{l+1}}...\partial_{\tau_i}g),
\end{split}
\end{equation}
where $G^{\tau_1...\tau_i}$ is fully-symmetric over its indices and $f, g$ are functions of Minkowski spacetime coordinates.


\end{document}